\documentclass[global,twocolumn]{svjour}
\usepackage{latexsym}
\usepackage{graphics}
\journalname{Applied physics B}
\begin{document}
\title{Time interval distributions of atoms in atomic beams}
\author{Michael K\"ohl, Anton \"Ottl, Stephan Ritter, Tobias Donner, Thomas
Bourdel \and Tilman Esslinger}

\institute{Institute for Quantum Electronics, ETH Z\"urich,
CH-8093 Z\"urich, Switzerland}

\date{Received: \today / Revised version: date}
%\pacs{03.75.Pp, 05.30.Jp, 07.77.Gx, 42.50.Pq}
\authorrunning{Michael K\"ohl at al.}
 \maketitle

\begin{abstract}
We report on the experimental investigation of two-particle
correlations between neutral atoms in a Hanbury Brown and Twiss
experiment. Both an atom laser beam and a pseudo-thermal atomic
beam are extracted from a Bose-Einstein condensate and the atom
flux is measured with a single atom counter. We determine the
conditional and the unconditional detection probabilities for the
atoms in the beam and find good agreement with the theoretical
predictions.
\end{abstract}

\section{Introduction}

Hanbury Brown and Twiss experiments \cite{hanbury1956} play a
central role in investigating correlations between identical
particles. The underlying idea to these experiments is that
intensity fluctuations and their correlations are tightly linked
to the quantum mechanical state of a system. The statistical
properties of a beam of photons or atoms can be accessed in
counting experiments. A famous example of how the statistics vary
for different quantum states of light is the distinct difference
of photon correlations in thermal light beams and laser beams
\cite{Arecchi1965,arecchi1966c}.

For ideal bosonic atoms the same quantum statistical properties
are expected as for the case of photons. First experiments
investigating the second order correlation function of
laser-cooled thermal atoms have already been undertaken several
years ago and the bunching behavior could be observed
\cite{yasuda1996}. The availability of quantum degenerate atomic
gases \cite{anderson1995} has added a new dimension since now even
coherent states of matter can be investigated. Recently, the
second order correlation function $g^{(2)}(\tau)$ of quantum
degenerate atoms has been observed \cite{ottl2005,schellekens2005}
and the feasibility of studying Hanbury Brown and Twiss
correlations of atoms released from an optical lattice has been
demonstrated \cite{altman2004,folling2005}. Moreover, atom-atom
correlations have also been observed in the dissociation process
of ultracold molecules \cite{greiner2005}.

The second order temporal correlation function of the stationary
field $\psi$
\begin{equation}
g^{(2)}(\tau)= \frac{\langle \psi^\dag(\tau) \psi^\dag(0) \psi(0)
\psi(\tau)\rangle}{\langle \psi^\dag(0) \psi(0)
\rangle^2}
\end{equation}
reveals valuable information about the intensity noise and the
two-particle correlations in the sample. In particular, for an
atomic beam released from a Bose-Einstein condensate the
correlation function was found to be equal to unity revealing the
second order coherence of the atom beam \cite{ottl2005}. Together
with the measurement of first order coherence of this beam
\cite{kohl2001a} this showed that atomic beams extracted from a
Bose-Einstein condensate indeed are the matter wave analogue of an
optical laser.

In this manuscript we discuss the two-particle correlations of
atoms in an atomic beam extracted from a Bose-Einstein condensate.
We measure the conditional and the unconditional probabilities for
atom detection which constitutes a complementary view on
two-particle correlations as compared to an analysis of the second
order correlation function.

\section{Experimental setup}
Our experimental setup combines the techniques for the production
of atomic Bose-Einstein condensates with single atom detection by
means of an ultrahigh finesse optical cavity
\cite{ottl2005,ottl2006}. We collect $10^9$ $^{87}$Rb atoms in a
vapor cell magneto-optical trap which is loaded from a pulsed
dispenser source. After polarization gradient cooling and optical
pumping into the $|F=1, m_F=-1\rangle$ hyperfine ground state we
magnetically transfer the atoms over a distance of 8\,cm into a
magnetic QUIC trap \cite{esslinger1998}. In this magnetic trap we
perform radio frequency induced evaporative cooling of the atomic
cloud and obtain almost pure Bose-Einstein condensates of $2\times
10^6$ atoms. After evaporation we relax the confinement of the
atoms to the final trapping frequencies
$(\omega_x,\omega_y,\omega_z)= 2 \pi \times (39,7,29)$\,Hz, where
$z$ denotes the vertical axis.

For output coupling an atom laser beam we apply a weak continuous
microwave  field to locally spin flip atoms inside the
Bose-Einstein condensate into the $|F=2, m_F=0\rangle$ state.
These atoms do not experience the magnetic trapping potential but
are released from the trap and form a well collimated beam which
propagates downwards due to gravity \cite{Bloch1999}. The output
coupling is performed near the center of the Bose condensate for a
duration of 500\,ms during which we extract about $3\times 10^3$
atoms.

Alternatively, we create a beam with pseudo-thermal correlations
from a Bose-Einstein condensate. This is in close analogy to
changing the coherence properties of a laser beam by means of a
rotating ground glass disc \cite{martienssen1964,Arecchi1965}.
Instead of applying a monochromatic microwave field for output
coupling we use a broadband microwave field. We employ a white
noise radio-frequency generator in combination with a quartz
crystal band pass filter which sets the bandwidth $\Delta f$ of
the noise. The filter operates at a frequency of a few MHz and the
noise signal is subsequently mixed to a fixed frequency signal
close to the hyperfine transition at 6.8\,GHz to match the output
coupling frequency.

\begin{figure}[htbp]
\resizebox{\columnwidth}{!}{\includegraphics{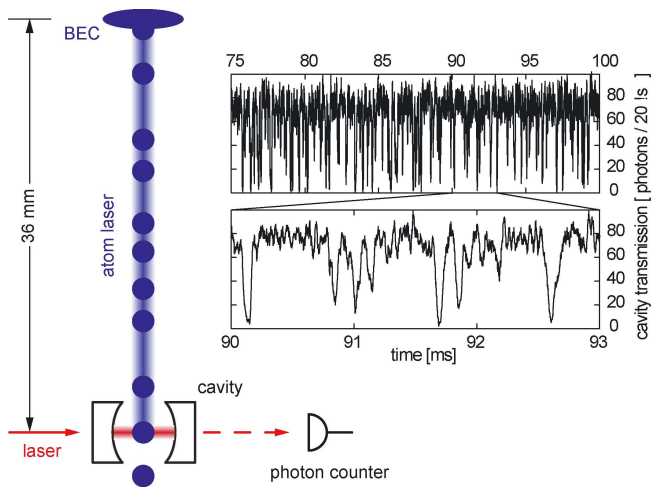}}
\caption{Schematic view of the experimental setup. A Bose-Einstein
condensate (BEC) is produced 36\,mm above an ultrahigh finesse
optical cavity. Using microwave output coupling we extract an
atomic beam from the condensate which passes through the cavity.
The intensity of a laser beam probing the cavity is modified by
the presence of a single atom inside the cavity resulting in the
characteristic dips of the cavity transmission.} \label{fig1}
\end{figure}

After propagating a distance of 36\,mm the atoms enter the
ultrahigh finesse optical cavity (see Fig. \ref{fig1}). The cavity
consists of two identical mirrors separated by 178\,$\mu$m. Their
radius of curvature is 77.5\,mm resulting in a gaussian TEM$_{00}$
mode with a waist of $w_0=26\,\mu$m. The coupling strength between
a single Rb atom and the cavity field is $g_0=2 \pi \times
10.4$\,MHz on the $F=2 \rightarrow F^\prime=3$ transition of the
D$_2$ line. The cavity has a finesse of $3\times 10^5$ and the
decay rate of the cavity field is $\kappa= 2 \pi \times 1.4$\,MHz.
The spontaneous decay rate of the atomic dipole moment is $\gamma=
2 \pi \times 3$\,MHz. Since $g_0 \gg \kappa, \gamma$ we operate in
the strong coupling regime of cavity QED. The cavity resonance
frequency is stabilized by means of a far-detuned laser with a
wavelength of 830\,nm using a Pound-Drever-Hall locking scheme.

The cavity is probed by a weak, near resonant laser beam whose
transmission is monitored by a single photon counting module. We
find a shot-noise limited transmission of photons through the
empty cavity. The presence of an atom inside the cavity results in
a drop of the transmission. The intensity and the frequency of the
detection laser are optimized to yield a maximum detection
efficiency for the released atoms which is $(23\pm 5)\%$
\cite{ottl2006}. This number is primarily limited by the size of
the atom laser beam which exceeds the cavity mode cross section.
The atoms enter the cavity with a velocity of 84\,cm/s. The
resulting dead time of our detector of approximately 70\,$\mu$s is
short compared to the time scale of the correlations, which allows
us to perform Hanbury Brown and Twiss type measurements with a
single detector \cite{Mandel1965}.

We record the cavity transmission for the period of atom laser
operation (typically 0.5\,s) and average the photon counting data
over 20\,$\mu$s. Using a peak detection routine we determine the
arrival time of atoms in the cavity, requiring that the cavity
transmission drops below the empty cavity value by at least four
times the standard deviation of the photon shot noise.

\section{Results}

We first investigate the distribution function of the time
intervals between successive atom counting events. This represents
a "start-stop" measurement, where a time counter is triggered by
an atom detection event and stopped by the next detection event
\cite{arecchi1966c}. From the histogram of the measured time
intervals we obtain the conditional probability $p(t|t+\tau)$ of
detecting the {\em next} atom a time $\tau$ later than an initial
atom observed at $t$. These exclusive pair correlations, for which
we restrict ourselves to consecutive atom detection events, are
distinguished from the non-exclusive pair correlations measured by
the second order correlation function $g^{(2)}(\tau)$. There the
pairwise time differences between {\em all} atoms are evaluated.

For an average count rate $\nu$ the conditional detection
probability density for a coherent beam is given by
\cite{glauber1972}
\begin{equation}
p_{coh}(0|\tau) = \nu \exp(-\nu \tau). \label{eqn2}
\end{equation}
In contrast, for a thermal state of bosons one finds
\cite{glauber1972}
\begin{equation}
p_{th}(0|\tau) = \frac{2 \nu}{(1+\nu \tau)^3}. \label{eqn3}
\end{equation}
For $\tau = 0$ the thermal probability density is twice as large
as the coherent probability density. This reflects the increased
thermal fluctuations and the bunching behavior in pair
correlations of bosonic particles.
\begin{figure}[htbp]
\resizebox{\columnwidth}{!}{\includegraphics{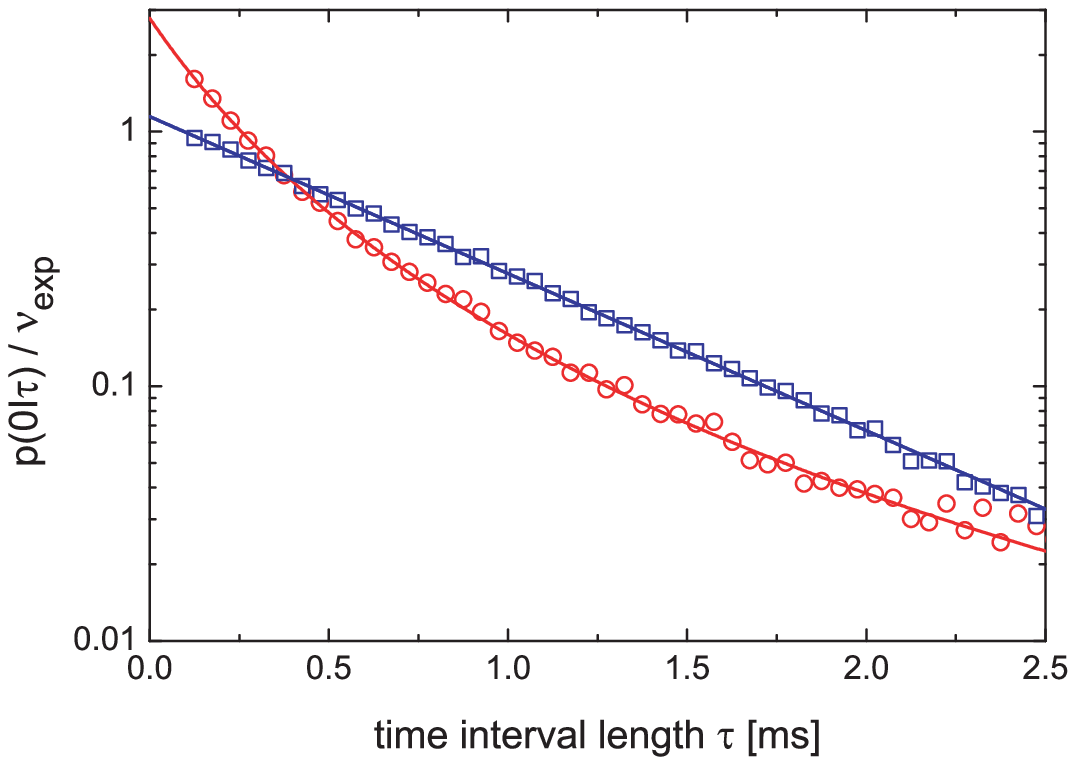}}
\caption{Conditional detection probability $p(0|\tau)$. The
frequency distribution of the time intervals between two
successive atom counts is shown for the atom laser (squares) and
for a pseudo-thermal beam with a bandwidth of $\Delta f=90\,$Hz
(circles). The lines are fits according to equations (2) and (3).}
\label{fig2}
\end{figure}

In Fig. \ref{fig2} we compare our data with this theory. For the
pseudo-thermal atomic beam we have chosen a bandwidth of $\Delta
f=90$\,Hz and analyzed time intervals short compared to the
coherence time $\tau_c=1/\Delta f=11$\,ms. We normalize the
measured probability by the measured count rate
$\nu_{exp}=1.5\times 10^{3}$\,s$^{-1}$ and fit the result with the
functions given in equation\,(\ref{eqn2}) and
equation\,(\ref{eqn3}) allowing for some overall scaling factor.
From the fits we obtain the average count rates $\nu=1.4\times
10^{3}$\,s$^{-1}$ and $\nu=1.6\times 10^{3}$\,s$^{-1}$ for the
atom laser beam and the pseudo thermal beam, respectively, which
compares well with the experimentally determined flux $\nu_{exp}$
for both cases. For $\tau=0$ we find that the data for the atom
laser beam exceed $p(0|\tau)/\nu_{exp}=1$ by approximately $15\%$.
Similarly, the results for the pseudo-thermal beam exceed
$p(0|\tau)/\nu_{exp}=2$ by approximately $30\%$. This could be
attributed to the dead time of our detector -- which is about
70$\,\mu$s \cite{ottl2005} -- during which we cannot detect a
possible consecutive event. We estimate the probability for a
second atom arriving within the dead-time of the detector to be
5$\%$ for the atom laser beam and 10$\%$ for atoms in the
pseudo-thermal beam. With this probability a later atom might
falsely be identified being consecutive to the initial event which
overestimates the number of time intervals larger than the
detector dead time. Moreover, the experimental count rate is
underestimated by the same factor contributing also to the
enhancement of the data above the theoretical expectation.
\begin{figure}[htbp]
\resizebox{\columnwidth}{!}{\includegraphics{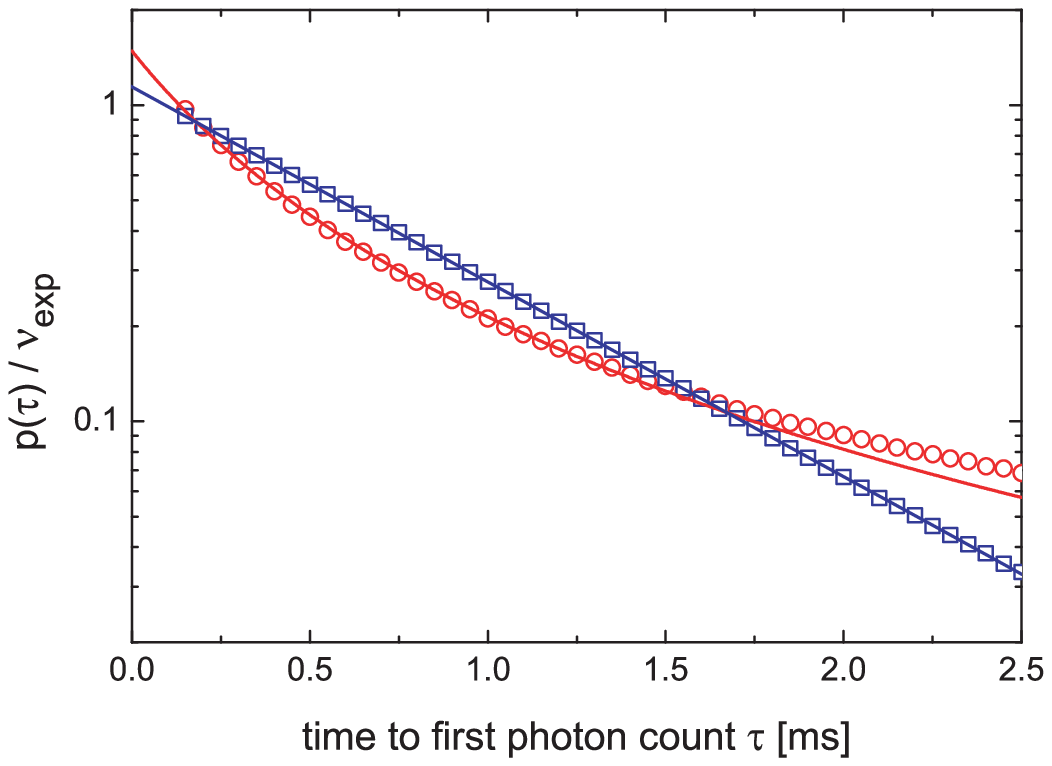}}
\caption{Unconditional detection probability $p(\tau)$. The
frequency distribution of the length of intervals between a
randomly chosen start point and the subsequent atom detection
event is shown for the atom laser (squares) and for a
pseudo-thermal beam with a bandwidth of $\Delta f=90$\,Hz
(circles). The lines a fits according to equations (4) and (5).}
\label{fig3}
\end{figure}

Next we study the unconditional probability of a single atom
detection event. The unconditional probability assumes that the
timer is started at a randomly chosen time and records the time to
the next atom detection event. For a coherent beam of atoms the
unconditional probability for a detection event $p(t)$ is equal to
the conditional probability investigated above \cite{glauber1972}
\begin{equation}
p_{coh}(\tau) = \nu \exp(-\nu \tau).
\end{equation}
This reflects the absence of any density correlations in a
coherent atomic beam. For a thermal state one finds
\begin{equation}
p_{th}(\tau) = \frac{\nu}{(1+\nu \tau)^2},
\end{equation}
which for $\tau=0$ differs from the corresponding conditional
probability by a factor of 2. The physical reason for this
difference lies in the bunching of thermal bosons, which enhances
the detection probability only for two nearby events measured by
the conditional probability. The unconditional probability
measures a single particle property and does not reveal a bunching
effect. In Fig. \ref{fig3} we show our measurements of the atom
detection probability for a randomly chosen initial start point
and find good agreement with the theoretical prediction. Similarly
to the results for the conditional probability we observe that the
experimental data for $\tau=0$ are larger than the theoretically
expected result of $p(\tau)/\nu_{exp}=1$ by the same relative
amount as in Fig.\,\ref{fig2}. We attribute this again to the dead
time of our detector as discussed above. The apparently better
data quality of Fig.\,\ref{fig3} as compared to Fig.\,\ref{fig2}
is due to the larger number of available time intervals for the
unconditional probability.

\section{Conclusion}
We have studied the interval distribution of atom detection in an
atom laser beam and a pseudo thermal atomic beam. We have
investigated both the conditional and the unconditional detection
probability and find good agreement with the theoretical
predictions. This complements the measurement of the second order
correlation function of the atomic beams \cite{ottl2005}.

\bigskip

We acknowledge stimulating discussions with F.T. Arecchi and  F.
Brennecke, and thank R. Glauber for suggesting this experiment.
This work is supported by SNF, QSIT, and the EU project OLAQUI.

\end{document}